\documentclass[aps,prl,floatfix,twocolumn]{revtex4}
\usepackage[dvips]{graphicx}
\usepackage[english]{babel}
\usepackage{amsmath}
\usepackage{amssymb}

\newcommand{\be}{\begin{equation}}
\newcommand{\ee}{\end{equation}}

\begin{document}

\title{Strongly interacting Fermi gases with density imbalance}
\author{J. Kinnunen, L. M. Jensen and P. T\"orm\"a}
\affiliation{Nanoscience Center, Department of Physics, P.O.Box 35, FIN-40014 University of 
Jyv\"askyl\"a, Finland}

\begin{abstract}
We consider density-imbalanced Fermi gases of atoms in the strongly
interacting, i.e. unitarity, regime. The Bogoliubov-de Gennes equations for
a trapped superfluid are solved. They take into account the finite
size of the system, as well as give rise to both phase separation and
Fulde-Ferrel-Larkin-Ovchinnikov (FFLO)-type oscillations in the order parameter. We show how
radio-frequency spectroscopy
reflects the phase separation, and can provide direct evidence of the
FFLO-type oscillations via observing the nodes of the order parameter.  
\end{abstract}

\maketitle
Bardeen-Cooper-Schrieffer (BCS) pairing is behind various forms of
superconductivity and superfluidity. A prerequisite for BCS pairing is the matching of
the Fermi energies of the two pairing components (e.g.\ spin up and spin down
electrons). In the case of spin-imbalanced (polarized) Fermi energies, non-standard forms of pairing are predicted, 
such as Fulde-Ferrel-Larkin-Ovchinnikov (FFLO) pairing~\cite{Fulde1964,Larkin1965}, 
interior gap superfluidity or breached pairing 
\cite{Sarma1963,Combescot2001,Liu2003,Deb2004,Gubankova2003,Liao2003} and phase separation \cite{Bedaque2003,Sheehy2006}.  
These exotic quantum states have relevance to many fields of
physics, e.g.\ superconductors in a magnetic field, neutron-proton pairing in nuclear matter, and color superconductivity in high density QCD; for a
recent extensive review see~\cite{Casalbuoni2004}.
The newly realized strongly interacting superfluid Fermi gases
\cite{Jochim2003b,Greiner2003b,Regal2004a,Zwierlein2004a,Bartenstein2004b,Kinast2004a,Chin2004,Kinast2005a,Zwierlein2005b} offer a promising playground for the study of pairing and
superfluidity with
variable initial conditions --- also imbalanced Fermi energies. Indeed, first such experiments have recently been done, showing disappearance of
superfluidity and vortices with increasing spin-density imbalance
\cite{Zwierlein2005c}, and phase separation \cite{Zwierlein2005c,Partridge2005}.
Here we consider theoretically the density imbalanced Fermi gas in the
unitarity
regime, discuss the role of finite size effects, and show how phase separation and FFLO-type oscillations appear and can be observed in 
the RF-spectrum of the system. 

Atomic Fermi gases are confined in magnetic or optical traps and the
harmonic trapping
potential causes significant finite size
effects. For instance, for the density imbalanced system, clear phase separation of the majority
component, i.e. the component with the most atoms, towards the
edges of the trap has been experimentally observed in \cite{Zwierlein2005c,Partridge2005}.
Theoretically, the finite size can be taken into account by solving
the Bogoliubov-deGennes (BdG) equations in the trap geometry. This has
been done for the density imbalanced case in the BCS-limit (weak coupling limit)
in \cite{Castorina2005,Mizushima2005a,Yang2005}. In
the opposite limit where the coupling is so strong that dimers
are formed and undergo Bose-Einstein condensation, the system has been described by
a mean-field treatment of a bosonic condensate interacting with
fermions in normal state within the local density approximation
\cite{Pieri2005c}. Here we consider the intermediate case, where the
interactions are
strong, but the pairing is still fermionic in nature. In the ultracold
atomic Fermi gases this corresponds to the Feshbach-resonance, or
unitarity, regime. This regime was considered both in~\cite{Zwierlein2005c,Partridge2005}, thereby our results provide direct comparison to the
experiments. 

We use the single-channel mean-field resonance superfluidity Hamiltonian~\cite{Bruun1999a, Ohashi2005b}
\be
\begin{split}
  H &= \sum_\sigma \int d{\bf r} \, \Psi_\sigma^\dagger ({\bf r}) \left[ -\frac{\nabla^2}{2m} + V_\mathrm{trap} (r) - \mu_\sigma \right] \Psi_\sigma
({\bf r}) \\
  &- U\int d{\bf r} \, \Psi_\uparrow^\dagger ({\bf r})  \Psi_\downarrow^\dagger ({\bf r}) \Psi_\downarrow ({\bf r})  \Psi_\uparrow ({\bf r}),
\end{split}
\ee
where $V_\mathrm{trap} = \frac{1}{2} m \omega_0^2 r^2$ is the
spherically symmetric harmonic trapping potential with $m$ the atom
mass and $\omega_0$ the trap oscillator frequency, $\mu_\sigma$ is the
chemical potential for atoms in hyperfine state $\sigma$, and $U$ is the
effective interaction strength. This describes a two-component ($\sigma = \uparrow , \downarrow$, now two different hyperfine states of the atom) gas
with s-wave contact interactions, where the strength of the interaction is tunable via a Feshbach resonance.

Following the treatment in Ref.~\cite{Ohashi2005b}, we expand the field operators in eigenstates of the harmonic potential
\be
  \Psi_\sigma ({\bf r}) = \sum_{nlm} \psi_{nlm} ({\bf r}) c_{nlm
    \sigma} = \sum_{nlm} R_{nl} (r) Y_{lm} (\hat r) c_{nlm \sigma} \nonumber
\ee
where $Y_{lm} (\hat r)$ are the standard spherical harmonics and the
radial part is given by 
\be
  R_{nl} (r) = \sqrt{2} \left( m\omega_0 \right)^{3/4} \sqrt{\frac{n!}{(n+1+1/2)!}} e^{-\frac{\overline{r}^2}{2}} \overline{r}^l L_n^{l+1/2} \left(
\overline{r}^2 \right), \nonumber
\ee
where $\overline{r} = \sqrt{m\omega_0} r$ and $L_n^{l+1/2}$ is the
Laguerre polynomial. This gives the Hamiltonian
\be
\begin{split}
  H = &\sum_{nlm\sigma} \xi_{nl\sigma} c_{nlm\sigma}^\dagger
  c_{nlm\sigma} - \frac{U}{2} \sum_{nn'lm\sigma} J_{nn'\sigma}^l
  c_{nlm\sigma}^\dagger c_{n'lm\sigma} \\
  &- \sum_{nn'lm} F_{nn'}^l \left[
    c_{nlm\uparrow}^\dagger c_{n'l-m\downarrow}^\dagger +
    h.c. \right], \label{eq:ham2}
\end{split}
\ee
where the single-particle energies are $\xi_{nl\sigma} = \hbar
\omega_0 (2n+l+3/2) - \mu_\sigma$, the matrix element
$
  F_{nn'}^l \equiv \int_0^\infty dr\, r^2 R_{nl} (r) \tilde \Delta (r) R_{n'l} (r)
$
corresponds to anomalous term that describes the Cooper pair field $\Delta (r)$ and
$
  J_{nn'\sigma}^l \equiv \sum_{\sigma' \neq \sigma} \int_0^\infty dr\, r^2 R_{nl} (r)
  n_{\sigma'} (r) R_{n'l} (r)
$
is the Hartree interaction term.
The order parameter is given by
\be
  \Delta (r) = U \sum_{nn'l} \frac{2l+1}{4\pi} R_{nl} (r)
  R_{n'l} (r) \langle c_{nl0\downarrow} c_{n'l0\uparrow}
  \rangle,
\label{eq:gapeq}
\ee
and the fermion densities are
\be
  n_\sigma (r) = \sum_{nn'l} \frac{2l+1}{4\pi}
  R_{nl} (r) R_{n'l} (r) \langle c_{nl0\sigma}^\dagger c_{n'l0 \sigma} \rangle.
\label{eq:denseq}
\ee
These equations are solved self-consistently for 
fixed atom numbers $N_\uparrow$ and $N_\downarrow$ by varying the
corresponding chemical potentials $\mu_\sigma$.
The Hamiltonian~(\ref{eq:ham2}) is diagonalised using the Bogoliubov
transformation and the resulting eigenstates are used to calculate the
excitation gap and density profiles from Eqs.~(\ref{eq:gapeq})
and~(\ref{eq:denseq}). The diagonalisation gives rise to positive and negative
eigenenergies $E_{nl\sigma}$, and since in general setting the
particle-hole symmetry is broken, we need to keep all the solutions.

\begin{figure}
  \centering
  \includegraphics[height=9cm]{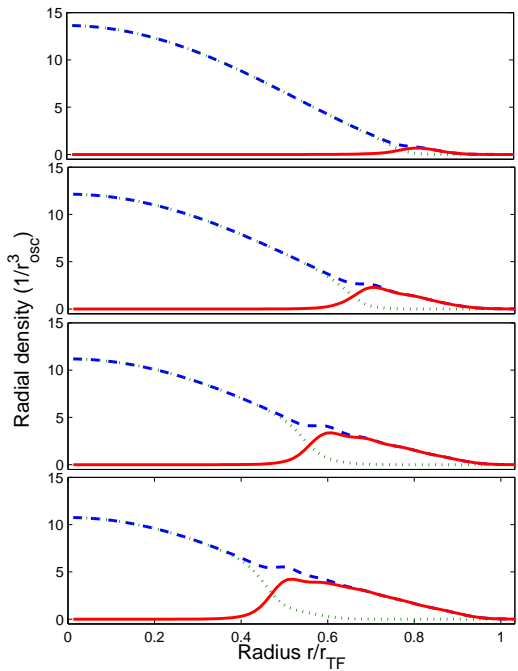}
  \caption{The radial density profiles for the majority ($\uparrow$-state, shown
    in dashed line) and minority ($\downarrow$-state, shown in dotted
    line) components $n_\sigma (r)$. Solid line shows the density difference
    as a function of distance from the center of the trap. The
    polarizations are $P = 0.04$ (upper panel), $P = 0.17$ (upper middle
    panel), $P = 0.34$ (lower middle panel), and $P = 0.49$ (lower panel).}
\label{fig:profiles}
\end{figure}

\begin{figure}
  \centering
  \includegraphics[height=7cm]{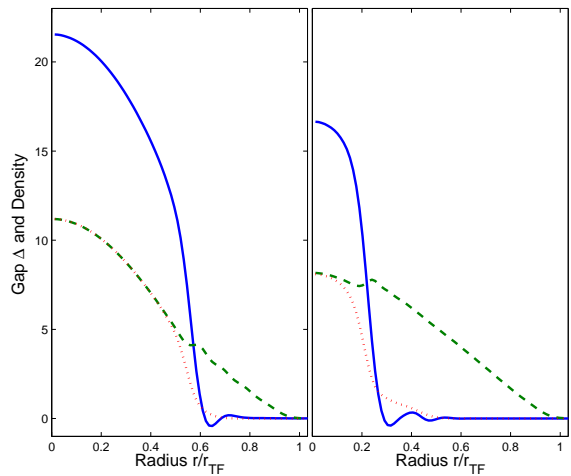}
  \caption{The radial density (dashed for $\uparrow$-state and dotted
    for $\downarrow$-state) and the gap (solid
    line) profiles for polarizations $P = 0.34$ (left) and $P = 0.88$ (right).}
\label{fig:singleprofile}
\end{figure}

We have solved the excitation gap $\Delta(r)$ and the density profiles
$n_\sigma(r)$ at zero temperature for 4570
atoms in the majority component ($N_\uparrow$) while the number of atoms
in the minority component ($N_\downarrow$) varies. We also tried
finite temperatures but observed no significant changes to the
picture. The parameters
have been chosen for $^6$Li in the unitarity limit with interaction
parameter $k_F a = -16$, where the Fermi wave vector is given by
$\frac{\hbar^2 k_F^2}{2m} = E_F = \hbar \omega_0 (6
N_\uparrow)^{1/3}$. The resulting profiles are shown in
Figs.~\ref{fig:profiles} and~\ref{fig:singleprofile} for several
polarizations $P \equiv \frac{N_\uparrow-N_\downarrow}{N_\uparrow+N_\downarrow}$.

\begin{figure}
  \centering
  \includegraphics[height=9cm]{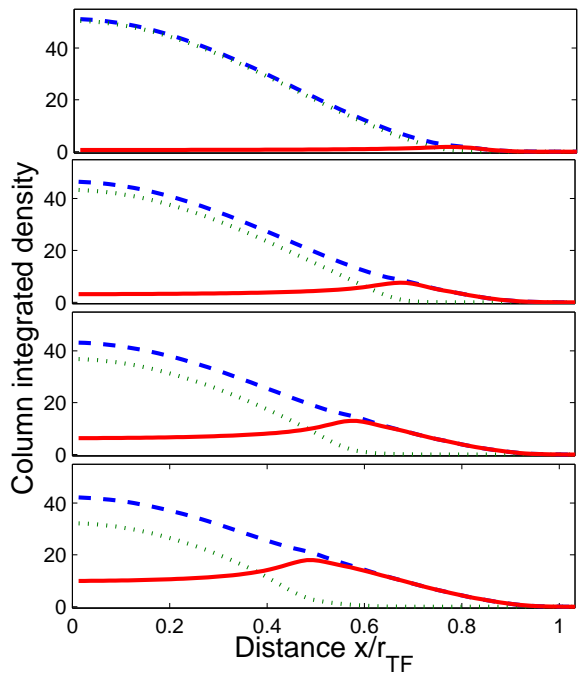}
  \caption{The column integrated density profiles for the majority (dashed) and minority
    (dotted) components. Solid line shows the density difference
    as a function of distance from the center of the trap. The
    polarizations are the same as in Fig.~\ref{fig:profiles}.}
\label{fig:axialprofiles}
\end{figure}

\begin{figure}
  \centering
  \includegraphics[height=6cm]{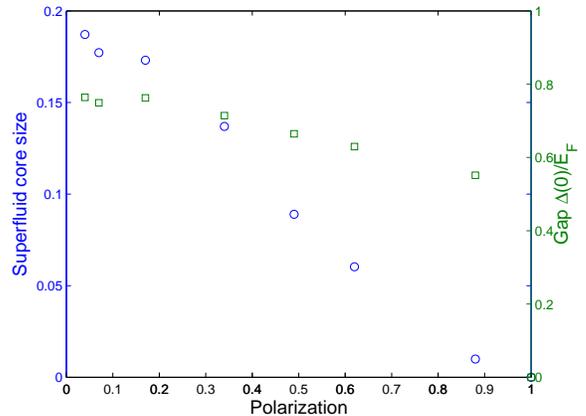}
  \caption{The value of the order parameter at the center of the trap (squares)
    and the fraction of the superfluid core as compared to the Fermi
    sphere of the major component (circles) plotted as functions of polarization $P$.}
\label{fig:gapwidth}
\end{figure}

\begin{figure}
  \centering
  \includegraphics[height=6cm]{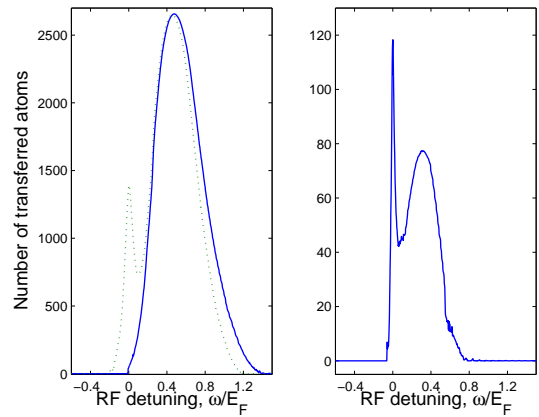}
  \caption{The RF-spectra for different components. Left panel shows
    the spectra for both majority (dotted) and minority (solid) components for polarization
    $P = 0.34$ and the right panel shows the spectrum for the minority
    component for polarization $P = 0.88$. The calculations are done at $T=0$}
\label{fig:spectra}
\end{figure}

The main qualitative features of the density profiles in
Figs.~\ref{fig:profiles} and~\ref{fig:axialprofiles} agree well
with the experiments in
Refs.~\cite{Zwierlein2005c,Partridge2005}. Fig.~\ref{fig:profiles}
shows the calculated radial density. Experimentally, the
densities are observed via imaging from one direction, leading to
column integrated densities. Corresponding integrated results from 
our calculations are shown in Fig.~\ref{fig:axialprofiles}. 
The density profiles show the phase separation into a superfluid core
and a normal fluid shell. The core corresponds to equal densities of
the two components while at the edges the majority component atoms
are dominating. In the column integrated picture
Fig.~\ref{fig:axialprofiles}, the excess amount of the majority
component at the edges of the trap leads to an effective density difference also at
$r=0$. This is seen both in Fig.~\ref{fig:axialprofiles} and in the
experiments~\cite{Zwierlein2005c,Partridge2005}. 
The bump in the density difference at the edge of the trap is an even more clear signature of phase separation. In~\cite{Partridge2005}, two regimes
were observed, namely: 
below polarization $P = 0.1$, a coexistence regime where the density difference does not show clear bumps (actually deviations from Thomas-Fermi
profiles were used as the
measure), and phase separation regime where these features are clearly visible. According to our results, there is no sharp transition between these
two
regimes; the phase separation starts immediately even for small polarizations, but the effect then may well be too small to observe. The absence
of a sharp transition does not exclude the possibility of a cross-over behaviour, where the amount of phase separation starts to grow
faster after a certain threshold. To this extent, we plot in Fig.~\ref{fig:gapwidth} the excitation gap $\Delta$ at center of the trap, as well as the
size (volume) of the superfluid core, as functions of the polarization. There might be a change
of slopes around the polarization $P=0.2$, but these results alone are not sufficient for any conclusive statements about a crossover behaviour.
However, they tell
clearly that the value of the gap at the center tends to stay quite constant, and the effect of the polarization is mainly to decrease the
superfluid core size. Indeed, the superfluid core size becomes negligible at some polarization $P$ between 0.6-0.9, agreeing well with the
experiments~\cite{Zwierlein2005c}, where $P=0.7$ was found to be a threshold for the disappearance of the condensate at the unitarity limit. 

In Fig.~\ref{fig:spectra} we show how the phase separation is reflected in
the RF-spectrum of the gas. RF-spectroscopy~\cite{Gupta2003a,Regal2003a} has been used e.g.~to observe the excitation gap of the
system~\cite{Torma2000,Chin2004,Kinnunen2004}. We calculate
the spectra using the method presented in~\cite{Kinnunen2005}. The results of  Fig.~\ref{fig:spectra} show a broad peak at finite RF-field detuning,
corresponding to paired atoms, and the detuning is related to the pairing energy $\Delta$. Note that there are equal amounts of
paired atoms for both components. The narrow peak at zero detuning corresponds to the nonpaired majority component atoms at the edge of the trap.
This could be a probe of phase separation, complementary to the straightforward observation of the density profiles, since it does not suffer from the
effects of column integration and provides a direct comparison between the amounts of the paired and nonpaired atoms. 
     
The gap and density profiles in Fig.~\ref{fig:singleprofile} show the oscillation of the order parameter and the density as a function of the
radial coordinate. Such oscillations have been
accounted~\cite{Castorina2005,Mizushima2005a,Yang2005} for a FFLO-type
phase. The FFLO-state in homogenous space leads
to oscillations of the order parameter, and is by definition pairing with unequal number of particles. In homogenous space, the FFLO pairing
starts only after a critical polarization. In our results,
oscillations are also visible for small polarizations, although
as tiny effects. This is understandable in the sense that, as the trap favours phase separation, the local polarization at the edges of
the trap becomes very easily of considerable size. Therefore locally one can fulfill the FFLO condition of exceeding a 
critical polarization, even when the total polarization is small. One could interpret the results in the following way: the trap tends to enforce a
normal BCS state at the
center of the trap and a FFLO-type state at the edges, and the significance of the latter grows with the total polarization. Is this FFLO-type state
observable? It may have existed in the experiments~\cite{Zwierlein2005c,Partridge2005}. The oscillations of the order parameter are accompanied by
oscillations of
the densities, and therefore, in principle, one could observe the FFLO characteristics from the density profiles. The column integration, however, tends
to wash out the oscillations as can be seen by comparing Fig.~\ref{fig:profiles} and Fig.~\ref{fig:axialprofiles}.
Thus, experimentally it may be difficult (although column integration can in principle be avoided by more advanced techniques) to detect the
FFLO-phase from the density profiles. The nodes of the order parameter, however, should be visible in the RF-spectrum of the minority component; they
produce a peak at zero detuning, reflecting a finite number of non-paired atoms also
in the minority component. This too, may be a small effect for some
parameters, e.g. in the left panel of Fig.~\ref{fig:spectra} such zero detuning peak is not visible in the minority
component. However, for parameter values that also produce a more prominent oscillation of the order parameter (Fig.~\ref{fig:singleprofile} right
panel), the zero detuning peak becomes clearly
visible, see the right panel in Fig.~\ref{fig:spectra}. This is a direct evidence for the nodes of the order parameter. Situations where such
signatures are large enough to be observed can
probably be achieved experimentally. For instance, we were restricted to spherical geometry due to computational reasons, but a cigar-shaped system is
likely to display more prominent oscillations.

In summary, we considered trapped, strongly interacting Fermi gases with unequal populations of the pairing components. We relate our findings to
recent experiments and suggest new ways of observing the phase separation and, especially, FFLO features. The system seems to be suited for detailed
studies of exotic forms of fermion pairing. Our results show that the trapping potential
affects the system in an essential way; spatial regions with different pairing characteristics tend to form and finite size effects have to be carefully taken
into account in understanding the system.    

{\it Acknowledgements} This project was supported by Academy of Finland and EUROHORCs (EURYI award,
Academy project numbers 106299, 205470), and the QUPRODIS project of
EU.



\bibliographystyle{apsrev}
\bibliography{./breach}

\end{document}